\documentclass[12pt]{iopart}
\usepackage{iopams}
\usepackage{amssymb}
\usepackage{graphicx,subfigure,epstopdf}

\usepackage{cite}
\eqnobysec
\usepackage[colorlinks,linktocpage,linkcolor=blue]{hyperref}

\begin{document}

\title[]{Models for characterizing the transition among anomalous diffusions with different diffusion exponents}

\author{Trifce Sandev$^{1,2,3}$, Weihua Deng$^{4}$ and Pengbo Xu$^{4}$}

\address{$^1$Radiation Safety Directorate, Partizanski odredi 143, P.O. Box 22, 1020 Skopje, Macedonia\\
$^2$Institute of Physics, Faculty of Natural Sciences and Mathematics, Ss. Cyril and Methodius University, P.O. Box 162, 1001 Skopje, Macedonia\\
$^3$Research Center for Computer Science and Information Technologies, Macedonian Academy of Sciences and Arts, Bul. Krste Misirkov 2, 1000 Skopje, Macedonia\\
$^4$School of Mathematics and Statistics, Gansu Key Laboratory of Applied Mathematics and Complex Systems, Lanzhou University, Lanzhou 730000, P.R. China}

\ead{trifce.sandev@drs.gov.mk, dengwh@lzu.edu.cn, and xupb09@lzu.edu.cn}
\vspace{10pt}
%\begin{indented}
%%\item[]February 2014
%\end{indented}

\begin{abstract}
Based on the theory of continuous time random walks (CTRW), we build the models of characterizing the transitions among anomalous diffusions with different diffusion exponents, often observed in natural world. In the CTRW framework, we take the waiting time probability density function (PDF) as an infinite series in three parameter Mittag-Leffler functions. According to  the models, the mean squared displacement of the process is analytically obtained and numerically verified, in particular, the trend of its transition is shown; furthermore the stochastic representation of the process is presented and the positiveness of the PDF of the position of the particles is strictly proved. Finally, the fractional moments of the model are calculated,  and the analytical solutions of the model with external harmonic potential are obtained and some applications are proposed.\\


\noindent (Some figures may appear in colour only in the online journal)
\end{abstract}


\section{Introduction}

The continuous time random walk (CTRW) model introduced by Montroll and Weiss \cite{MW} and applied in description of stochastic transport in disordered solids by Scher and Lax \cite{lax}, has become a very useful theory for description of anomalous diffusive processes in complex media, where deviation of the mean squared displacement (MSD) $\left\langle x^2(t)\right\rangle\simeq t^{\alpha}$ from the linear Brownian scaling with time exists. The parameter $\alpha$ is the anomalous diffusion exponent and we distinguish subdiffusion where $0<\alpha<1$, normal diffusion if $\alpha=1$, and superdiffusion if $\alpha>1$ \cite{bouchaud,report}. Subdiffusive phenomena are observed, for example, in the charge carrier motion in amorphous semiconductors \cite{scher,schubert}, tracer chemical dispersion in groundwater \cite{scher2}, and in the motion of submicron probes in living biological cells \cite{lene}. Apart from the subdiffusion, superdiffusive processes are observed in weakly chaotic systems \cite{swinney}, turbulence \cite{turbu}, search processes \cite{search}, diffusion in porous structurally inhomogeneous media \cite{zhokh}, to name but a few. Because of the complexity of the transport media and the multiple properties of the particles, generally the type of diffusions changes with the evolution of time. This paper focuses on modelling the whole transition procedure.

For the diffusion limit, i.e., $t \rightarrow +\infty$, one may derive the fractional Fokker-Planck equation (FFPE) \cite{report,barkai2000,epl}, describing subdiffusion or superdiffusion with some striking properties; for example, the subdiffusive CTRW processes and therefore the motions governed by the Fokker-Planck equation are weakly non-ergodic, which means that the long time and ensemble averages of physical observables are different, in contrast to, e.g., Brownian motion or Langevin equation motion \cite{he2}. The decoupled subdiffusive CTRW model in the long time limit yields the time fractional Fokker-Planck equation \cite{report}. Modifications of this well known approach include distributed order fractional diffusion equations \cite{chechkin_1}, and noisy CTRW \cite{noisy_ctrw}. Usually, the fractional equation governs the probability density function (PDF) of non-Gaussian anomalous diffusion process. For the Gaussian anomalous process, the corresponding Fokker-Planck equation is classical one with time-dependent coefficient \cite{YaoChen2017}, which can describe subdiffusion or superdiffusion or even the transitions among different types of diffusions, depending on the evolution of the coefficient with time. If the density of such processes can be represented as a subordination-type integral with a Gaussian kernel, as well as the Gaussian kernel is completely characterized by the first and the second moments, included the correlation, the same is the resulting process \cite{pagnini3}. %Obviously, not like normal diffusion, the first and second moments are not sufficient to determine the microscopic mechanism of the stochastic process, i.e., for anomalous diffusions, two stochastic processes with the same first and second moments can be completely different.
The stochastic processes which will be modeled in this paper are non-Gaussian, moreover, the diffusion type changes with the time marching. So, some new operators are needed for the governing equation of the evolution of the PDF of positions of the particles. In this paper we employed the elegant subordination approach to analyze the corresponding solution. It is worth mentioning that the link between classical diffusion and fractional diffusion through a subordination-type integral has been studied by Wyss and Wyss \cite{wyss}, Mainardi et al. \cite{mainardi_et_al}, and Pagnini et al. \cite{pagnini1,pagnini2}.

This paper is organized as follows. In Section 2, we introduce the CTRW models, which describe the transitions of anomalous dynamics; the basic idea is to choose the multi-scale waiting time and/or jump length PDF(s), i.e., the PDFs obey different trends for short time/distance, long time/distance, and in-between time/distance, specifically given as an infinite series in three parameter Mittag-Leffler functions; based on the proposed CTRW model, the fractional diffusion equation with Prabhakar time fractional derivative is derived. In Section 3, we prove the non-negativity of the PDF by employing the subordination approach and the properties of the complete monotone and Bernstein functions. In Section 4, we calculate and analyze the MSD and fractional moments, and show that the considered model describes various transitions among anomalous diffusions with different anomalous diffusion exponents. Section 4 investigates the generalized Fokker-Planck equation, and obtains relaxation of modes, the analytical solution of the system with harmonic potential, the survival probability, and the first passage time density. We conclude the paper with some comments in Section 6.

\section{Transition among anomalous diffusions: CTRW description}

Here we build the CTRW model describing the transitions among anomalous diffusions with different diffusion exponents. It is well known that, in Fourier-Laplace space, the PDF of the positions of the particles satisfies \cite{report,scher}
\begin{equation}\label{CTRW general}
\tilde{\hat{W}}(k,s)=\frac{1-\hat{\psi}(s)}{s}\frac{\tilde{W}_{0}(k)}{1-\hat{\psi}(s)\tilde{\lambda}(k)},
\end{equation}
where $\hat{\psi}(s)$ is the Laplace transform of the waiting time PDF $\psi(t)$, i.e., $\hat{\psi}(s)=\mathcal{L}[\psi(t)]=\int_{0}^{\infty}\psi(t)e^{-st}\,dt$, $\tilde{\lambda}(k)$ is the Fourier transform of the jump length PDF $\lambda(x)$, $\tilde{\lambda}(k)=\mathcal{F}\left[\lambda(x)\right]=\int_{-\infty}^{\infty}\lambda(x)e^{-ikx}\,dx$, and the initial condition is of the form $W_{0}(x)=W(x,0)$ and  $\tilde{W}_{0}(k)=\mathcal{F}\left[W_{0}(x)\right]$. If take the distribution of jump lengths as a Gaussian with variance $\Sigma^{2}=2\sigma^2$, i.e., the jump length PDF $\tilde{\lambda}(k)=e^{-\sigma^{2}k^{2}}\simeq1-\sigma^{2}k^2$ \cite{report}, and the waiting time PDF as exponential distribution $\psi(t)=\tau^{-1}\exp(-t/\tau)$ with the finite mean waiting time $T=\int_0^{\infty}t\psi(t)\,dt$, being equal to unity, then Eq. (\ref{CTRW general}) leads to the PDF of classical Brownian motion $\tilde{\hat{W}}(k,s)=\frac{1}{s+\mathcal{K}k^2}$, i.e., $W(x,t)=\frac{1}{\sqrt{4\pi \mathcal{K}t}}\exp\left(-\frac{x^2}{4\mathcal{K}t}\right)$, where $\mathcal{K}=\sigma^{2}/\tau$ is the diffusion coefficient with physical dimension $[\mathcal{K}]=\mathrm{m}^{2}\mathrm{s}^{-1}$. Furthermore, for a scale-free waiting time PDF $\psi(t)\simeq\tau^{\mu}/t^{1+\mu}$ with $0<\mu<1$, for which the characteristic waiting time $T$ diverges, the CTRW theory Eq. (\ref{CTRW general}) yields the following PDF $\tilde{\hat{W}}(k,s)=\frac{s^{\mu-1}}{s^{\mu}+\mathcal{K}_{\mu}k^2}$, where $\mathcal{K}_{\mu}=\sigma^2/\tau^{\mu}$ is the generalized diffusion coefficient with physical dimension $[\mathcal{K}_{\mu}]=\mathrm{m}^{2}\mathrm{s}^{-\mu}$ \cite{report}. This PDF is a solution of the time fractional diffusion equation exhibiting monoscaling behavior\footnote{By applying the inverse Fourier-Laplace transform, one can show that the PDF $W(x,t)$ obeys fractional diffusion equation with Caputo time fractional derivative \cite{report}, i.e., ${^{\mathrm{C}}}\mathcal{D}_{0+}^{\mu}W(x,t)=\mathcal{K}_{\mu}\frac{\partial^{2}}{\partial x^{2}}W(x,t)$.}. Moreover, in \cite{sandev_fcaa2015} we introduced generalized waiting time PDF which recovers the previously mentioned cases of classical and fractional diffusion equation, as well as time fractional distributed order and tempered in time diffusion equations.

In this paper, we first consider the following waiting time PDF in Laplace space
\begin{equation}\label{psi(s) prabhakar}
\hat{\psi}(s)=\frac{1}{1+(s\tau)^{\mu}\left[1+(s\tau)^{-\rho}\right]^{\gamma}},
\end{equation}
where $0<\rho<\mu<1$, $0<\gamma<1$. To ensure the non-negativity of the waiting time PDF $\psi(t)$, its Laplace transform $\hat{\psi}(s)$ should be completely monotone function \cite{book_bernstein}\footnote{The function $g(s)$ is a completely monotone if $(-1)^{n}g^{(n)}(s)\geq0$ for all $n\geq0$ and $s>0$. Product of completely monotone functions is completely monotone function too. An example of completely monotone function is $s^{\alpha}$, where $\alpha<0$.}, which means that $1+(s\tau)^{\mu}\left[1+(s\tau)^{-\rho}\right]^{\gamma}$, i.e., $(s\tau)^{\mu}\left[1+(s\tau)^{-\rho}\right]^{\gamma}$, should be a Bernstein function \cite{berg}\footnote{A given function $f(s)$ is a Bernstein function if $(-1)^{n-1}f^{(n)}(s)\geq0$ for all $n\in N$ and $s>0$. An example of Bernstein function is $s^{\alpha}$, where $0<\alpha<1$. If $f(s)$ is a complete Bernstein function then $g(s)=1/f(s)$ is a completely monotonic function.}; it is a Bernstein function if $0<\mu/\gamma<1$ and $0<\mu/\gamma-\rho<1$ \cite{garra_amc}\footnote{The function $$(s\tau)^{\mu}\left[1+(s\tau)^{-\rho}\right]^{\gamma}=\left[(s\tau)^{\mu/\gamma}+(s\tau)^{\mu/\gamma-\rho}\right]^{\gamma}$$ is a Bernstein function. This is satisfied if $s^{\mu/\gamma}$ and $s^{\mu/\gamma-\rho}$ are Bernstein functions, since linear combination of Bernstein functions is a Bernstein function, and the composition $f_{1}\circ f_{2}$ of Bernstein functions $f_{1}$ and $f_{2}$ is again a Bernstein function.}. Inserting the waiting time PDF (\ref{psi(s) prabhakar}) and the jump length PDF of Gaussian form in the case of long wavelength approximation, i.e., $\tilde{\lambda}(k)\simeq1-\sigma^{2}k^{2}$, into (\ref{CTRW general}), we get
\begin{equation}\label{CTRW}
\tilde{\hat{W}}(k,s)=\frac{s^{\mu-1}\left[1+(s\tau)^{-\rho}\right]^{\gamma}}{s^{\mu}\left[1+(s\tau)^{-\rho}\right]^{\gamma}+\mathcal{K}_{\mu}k^{2}}\tilde{W}_{0}(k),
\end{equation}
where $\mathcal{K}_{\mu}=\sigma^{2}/\tau^{\mu}$, and the generalized diffusion coefficient has a dimension $\left[\sigma^{2}/\tau^{\mu}\right]=\mathrm{m}^{2}\mathrm{s}^{-\mu}$, as it should be. Here it can be noted that this equation is valid only in the long wavelength approximation \cite{barkai_cp2002}. We discuss this point later in the calculation of the fractional moments. After some rearrangements, we arrive at
\begin{eqnarray}\label{diff eq FL}
   s^{\mu}\left[1+(s\tau)^{-\rho}\right]^{\gamma}\tilde{\hat{W}}(k,s)  -s^{\mu-1}\left[1+(s\tau)^{-\rho}\right]^{\gamma}\tilde{W}_{0}(k)=-\mathcal{K}_{\mu}k^{2}\tilde{\hat{W}}(k,s).\nonumber\\
\end{eqnarray}

Before to continue with further calculations, we would like to introduce the regularized Prabhakar derivative ${^{\mathrm{C}}}\mathcal{D}_{\rho,\omega,0+}^{\gamma,\mu}$ with $0<\mu<1$, which has been shown to have applications in the fractional Poisson process, defined by \cite{garra_amc} (see also \cite{polito})
\begin{eqnarray}\label{prabhakar derivative}
{^{\mathrm{C}}}\mathcal{D}_{\rho,\nu,0+}^{\gamma,\mu}f(t)=\left(\mathcal{E}_{\rho,1-\mu,\nu,0+}^{-\gamma}\frac{d}{dt}f\right)(t),
\end{eqnarray}
where $\mu, \nu, \gamma, \rho \in C$, $\Re(\mu)>0$, $\Re(\rho)>0$. Here
\begin{eqnarray}\label{prabhakar integral}
\left(\mathcal{E}_{\rho,\mu,\nu,0+}^{\gamma}f\right)(t)=\int_{0}^{t}(t-t')^{\mu-1}E_{\rho,\mu}^{\gamma}\left(\nu(t-t')^{\rho}\right)f(t')\,dt'
\end{eqnarray}
is the Prabhakar integral \cite{prabhakar}, and
\begin{equation}\label{ML three}
E_{\rho,\mu}^{\gamma}(t)=\sum_{k=0}^{\infty}\frac{(\gamma)_k}{\Gamma(\rho k+\mu)}\frac{t^k}{k!},
\end{equation}
where $(\gamma)_k=\Gamma(\gamma+k)/\Gamma(\gamma)$ -- the Pochhammer symbol, is the three parameter Mittag-Leffler (M-L) function \cite{prabhakar} (two parameter M-L function appeared in the representation of the waiting time PDF for fractional diffusion equation in \cite{hilfer_and_anton}). For $\gamma=0$ the Prabhakar integral becomes the Riemann-Liouville (R-L) fractional integral \cite{hilfer_ book,podlubny}
\begin{eqnarray}\label{RL integral}
{^{\mathrm{RL}}}I_{0+}^{\mu}f(t)=\frac{1}{\Gamma(\mu)}\int_{0}^{t}(t-t')^{\mu-1}f(t')\,dt'.
\end{eqnarray}
The Prabhakar derivative has been used for the description of dielectric relaxation phenomena \cite{garrappa,garrappa_fcaa}, in the fractional Maxwell model of the linear viscoelasticity \cite{giusti}, in mathematical modeling of fractional differential filtration dynamics \cite{Bulavatsky}, as well as in the generalized Langevin equation modeling \cite{sandev_mathematics2017}. Its Laplace transform is given by \cite{garra_amc}
\begin{eqnarray}\label{Prabhakar derivative Laplace}
\mathcal{L}\left[{^{\mathrm{C}}}\mathcal{D}_{\rho,\nu,0+}^{\gamma,\mu}f(t)\right](s)= s^{\mu}\left(1-\nu s^{-\rho}\right)^{\gamma}\mathcal{L}\left[f(t)\right](s)\nonumber -s^{\mu-1}\left(1-\nu s^{-\rho}\right)^{\gamma}f(0+),
\end{eqnarray}
where $ \Re(s)>|\nu|^{1/\rho}$. This relation follows from the Laplace transform formula for the M-L function \cite{prabhakar}, i.e.,
\begin{eqnarray*}
% \nonumber % Remove numbering (before each equation)
\mathcal{L}\left[t^{\mu-1}E_{\rho,\mu}^{\gamma}(\nu t^{\rho})\right](s)=\frac{s^{\rho\gamma-\mu}}{\left(s^{\rho}-\nu\right)^{\gamma}},
\end{eqnarray*}
$\Re(s)>|\nu|^{1/\rho}$. Note that for $\gamma=0$, Prabhakar derivative corresponds to Caputo derivative with Laplace transform given by \cite{mainardi_book}
\begin{eqnarray*}
\mathcal{L}\left[{^{\mathrm{C}}}\mathcal{D}_{0+}^{\mu}f(t)\right](s)=s^{\mu}\mathcal{L}\left[f(t)\right](s)-s^{\mu-1}f(0+).
\end{eqnarray*}

Then we go back to consider Eq. (\ref{diff eq FL}). After performing the inverse Fourier-Laplace transform, we get the time fractional diffusion equation
\begin{eqnarray}\label{diff eq}
{^{\mathrm{C}}}\mathcal{D}_{\rho,-\nu,0+}^{\gamma,\mu}W(x,t)=\mathcal{K}_{\mu}\frac{\partial^{2}}{\partial x^{2}}W(x,t),
\end{eqnarray}
where $\nu=\tau^{-\rho}$, $\tau$ is a time parameter with physical dimension $[\tau]=\mathrm{s}$, $\mathcal{K}_{\mu}$ is the generalized diffusion coefficient with physical dimension $\left[\mathcal{K}_{\mu}\right]=\mathrm{m}^{2}\mathrm{s}^{-\mu}$, being easily shown by dimensional analysis of Eq.~(\ref{diff eq}). The initial condition is given by
\begin{eqnarray}\label{initial condition}
W(x,0+)=W_{0}(x),
\end{eqnarray}
and the boundary conditions for the PDF $W(x,t)$ and its derivative $\frac{\partial}{\partial x}W(x,t)$ are set to zero at infinity $x=\pm\infty$.

Let us now consider the transition of the waiting times. For the waiting time PDF $\psi(t)$ with Laplace transform (\ref{psi(s) prabhakar}), by using the series expansion approach \cite{podlubny}, we obtain
\begin{eqnarray}\label{psi(t)_prabhakar}
\psi(t)=\frac{1}{\tau}\sum_{n=0}^{\infty}(-1)^{n}\left(\frac{t}{\tau}\right)^{\mu n+\mu-1}E_{\rho,\mu n+\mu}^{\gamma n+\gamma}\left(-\left[\frac{t}{\tau}\right]^{\rho}\right).
\end{eqnarray}
Series in three parameter M-L functions of form (\ref{psi(t)_prabhakar}) are convergent \cite{sandev_physa} (for the detailed analysis of convergence of series in M-L functions, see \cite{paneva}). From the definition (\ref{ML three}), it follows that for small argument one has \cite{sandev_pre2015}
\begin{eqnarray}\label{ML three small argument}
E_{\alpha,\beta}^{\delta}(-t^{\alpha})\simeq\frac{1}{\Gamma(\beta)}-\delta\frac{
t^{\alpha}}{\Gamma(\alpha+\beta)}\simeq\frac{1}{\Gamma(\beta)}\exp\left(-\delta
\frac{\Gamma(\beta)}{\Gamma(\alpha+\beta)}t^{\alpha}\right),
\end{eqnarray}
thus, for the short time limit $t/\tau\ll1$, there exists
\begin{eqnarray}\label{psi(t)_prabhakar short}
\psi(t)\simeq\frac{1}{\tau}\left(\frac{t}{\tau}\right)^{\mu-1}E_{\mu,\mu}\left(-\left[\frac{t}{\tau}\right]^{\mu}\right)\simeq\frac{1}{\tau}\frac{(t/\tau)^{\mu-1}}{\Gamma(\mu)},
\end{eqnarray}
while for the long time limit $t/\tau\gg1$, it has
\begin{eqnarray}\label{psi(t)_prabhakar long}
\psi(t)&\simeq\frac{1}{\tau}\left(\frac{t}{\tau}\right)^{\mu-\rho\gamma-1}E_{\mu-\rho\gamma,\mu-\rho\gamma}\left(-\left[\frac{t}{\tau}\right]^{\mu-\rho\gamma}\right)\nonumber\\&\simeq\frac{\mu-\rho\gamma}{\tau}\frac{(t/\tau)^{-\mu+\rho\gamma-1}}{\Gamma(1-\mu+\rho\gamma)}.
\end{eqnarray}
Here we use the asymptotic expansion of the three parameter M-L function for $t\rightarrow\infty$ \cite{saxena,sandev_pre2015,sandev_fcaa2015,mainardi_garrappa,garrappa_fcaa,sandev_jmp,garra_garrappa}
\begin{eqnarray}\label{ml3 long}
E_{\alpha,\beta}^{\delta}(-t^{\alpha})=\frac{t^{-\alpha\delta}}{\Gamma(\delta)}\sum_{n=0}^{\infty}\frac{\Gamma(\delta+n)}{\Gamma(\beta-\alpha(\delta+n))}\frac{(-t^{\alpha})^{-n}}{n!}
\end{eqnarray}
for $0<\alpha<2$. Here we note that in this paper we provide the results in an exact form in terms of the M-L functions. However, for application purposes, we also use the previous asymptotic behaviors (\ref{ML three small argument}) and (\ref{ml3 long}) of these functions in terms of exponential and power-laws, in order to get the asymptotic limits of the considered models.

\begin{figure}
\centering{\resizebox{0.75\textwidth}{!}{\includegraphics{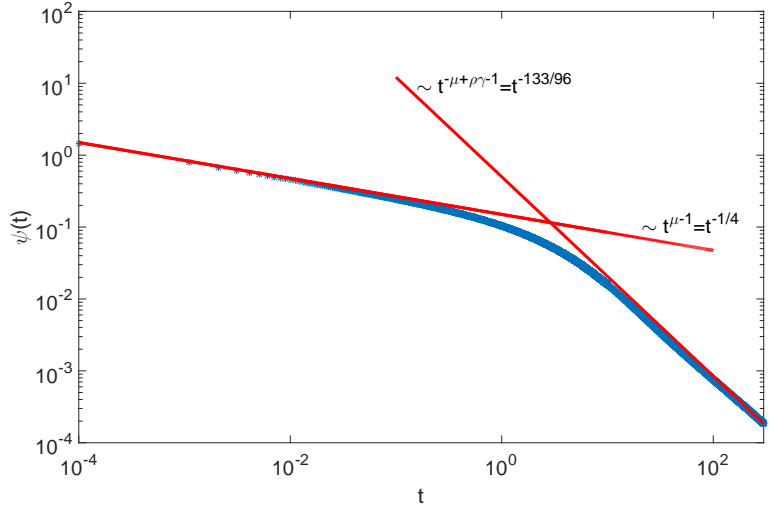}} \caption {Graphical representation of the  waiting time PDF with $\rho = 7/16$, $\mu = 3/4$, $\gamma = 5/6$, and $\tau = 10$; exact waiting time PDF (\ref{psi(t)_prabhakar}) (blue dot line), short time limit (\ref{psi(t)_prabhakar short}) (asymptotically behaves as $t^{\mu-1}$), long time limit (\ref{psi(t)_prabhakar long}) (asymptotically behaves as $t^{-\mu+\rho\gamma-1}$).}\label{fig_w}}
\end{figure}

From the above analysis, it can be noted that the parameters $\rho$ and $\gamma$ do not have influence on the particle behavior in the short time limit. In the long time limit all the parameters have influence on the diffusive behavior of the particle. This additionally will be confirmed below, by the analysis of the MSD. Graphical representation of the waiting time PDF is given in Fig.~\ref{fig_w}.

We can further consider a waiting time PDF $\psi(t)$ of the form
\begin{equation}\label{psi(s) tempered prabhakar}
\hat{\psi}(s)=\frac{1}{1+s\tau((s+b)\tau)^{\mu-1}\left[1+((s+b)\tau)^{-\rho}\right]^{\gamma}},
\end{equation}
where $b>0$ ($b$ has a dimension of inverse time, i.e., $[b]=\mathrm{s}^{-1}$) plays a role of tempering parameter, and all the other parameters are the same as the ones in Eq.~(\ref{psi(s) prabhakar}). Therefore, the model (\ref{psi(s) tempered prabhakar}) is a tempered model of the model described by Eq.~(\ref{psi(s) prabhakar}). For this waiting time PDF, in the long wavelength approximation, we obtain the diffusion equation
\begin{eqnarray}\label{tempered diff eq}
{^{\mathrm{TC}}}\mathcal{D}_{\rho,-\nu,0+}^{\gamma,\mu}W(x,t)=\mathcal{K}_{\mu}\frac{\partial^{2}}{\partial x^{2}}W(x,t),
\end{eqnarray}
where ${^{\mathrm{TC}}}\mathcal{D}_{\rho,-\nu,0+}^{\gamma,\mu}$ is the regularized Prabhakar derivative with tempering introduced in \cite{sandev_mathematics2017},
\begin{eqnarray}\label{tempered prabhakar derivative}
{^{\mathrm{TC}}}\mathcal{D}_{\rho,-\nu,0+}^{\gamma,\mu}f(t)=\left({^{\mathrm{T}}}\mathcal{E}_{\rho,1-\mu,-\nu,0+}^{-\gamma}\frac{d}{dt}f\right)(t)
\end{eqnarray}
with \cite{sandev_mathematics2017}
\begin{eqnarray}\label{tempered prabhakar integral}
\left({^{\mathrm{T}}}\mathcal{E}_{\rho,\mu,-\nu,0+}^{\gamma}f\right)(t) =\int_{0}^{t}e^{-b(t-t')}(t-t')^{\mu-1}\nonumber E_{\rho,\mu}^{\gamma}\left(-\nu[t-t']^{\rho}\right)f(t')\,dt'.
\end{eqnarray}

\begin{figure}
\centering{\resizebox{0.75\textwidth}{!}{\includegraphics{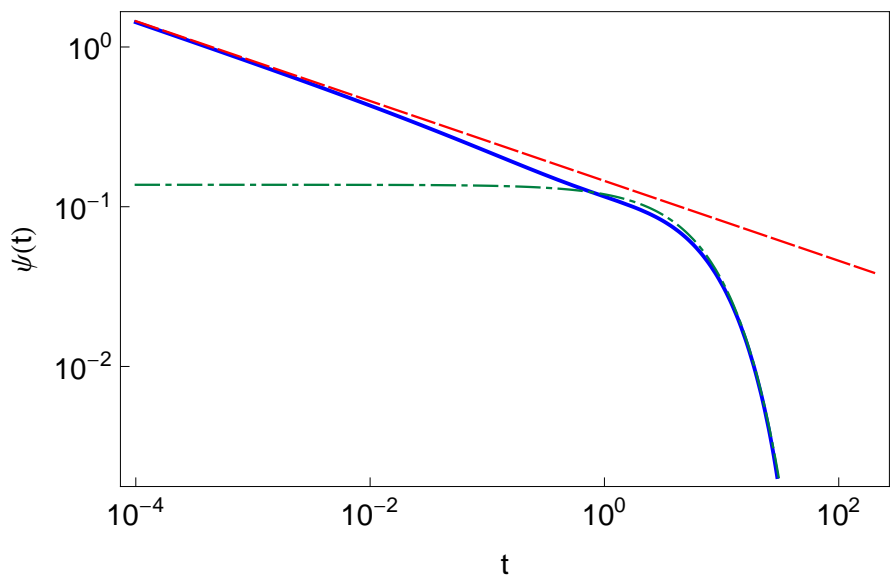}} \caption {Graphical representation of the waiting time PDF (\ref{psi(t)_tempered prabhakar}) for the same values of the parameters as those in Figure \ref{fig_w} ($\rho = 7/16$, $\mu = 3/4$, $\gamma = 5/6$, $\tau = 10$) and $b=1$ (blue solid line). The short time asymptotics (\ref{psi(t)_prabhakar short}) (red dashed line) and the long time exponential asymptotics (\ref{w PrT long time}) (green dot-dashed line) are in good agreement with the exact waiting time PDF.}\label{fig_w2}}
\end{figure}

\begin{figure}
\centering{\resizebox{0.75\textwidth}{!}{\includegraphics{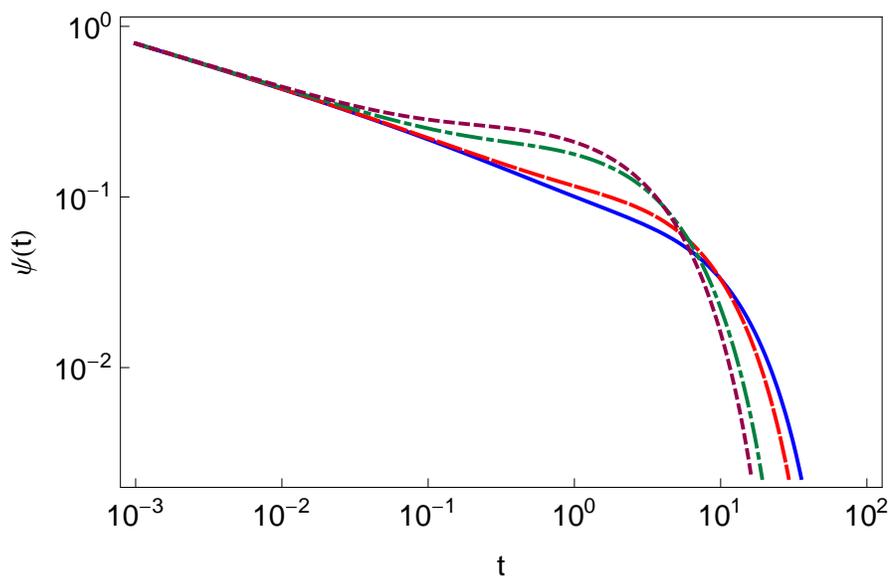}} \caption {Graphical representation of the  waiting time PDF (\ref{psi(t)_tempered prabhakar}) for $\rho = 7/16$, $\mu = 3/4$, $\gamma = 5/6$, $\tau = 10$ and $b=0.5$ (blue solid line), $b=1$ (red dashed line), $b=5$ (green dot-dashed line), $b=10$ (violet dotted line).}\label{fig_w2b}}
\end{figure}

For the waiting time PDF Eq.~(\ref{psi(s) tempered prabhakar}), there exists
\begin{eqnarray}\label{psi(t)_tempered prabhakar}
\fl\psi(t)=\frac{1}{\tau}\sum_{n=0}^{\infty}\frac{(-1)^{n}}{\tau^{n+1}}{^{\mathrm{RL}}}I_{0+}^{n+1}\left(e^{-bt}\left(\frac{t}{\tau}\right)^{(\mu-1)(n+1)-1} E_{\rho,(\mu-1)(n+1)}^{\gamma n+\gamma}\left(-\left[\frac{t}{\tau}\right]^{\rho}\right)\right),
\end{eqnarray}
where ${^{\mathrm{RL}}}I_{0+}^{\alpha}$ is the R-L integral (\ref{RL integral}). It can be seen that the waiting time PDF appears with exponential tempering. The short time limit shows the same behavior as the waiting time PDF (\ref{psi(t)_prabhakar}), i.e., (\ref{psi(t)_prabhakar short}), since the exponential tempering is negligible for small $t$, and for the long time limit, due to the exponential tempering, it yields exponential waiting time PDF
\begin{eqnarray}\label{w PrT long time}
\psi(t)=\frac{1}{\tau^{\ast}}\exp\left(-t/\tau^{\ast}\right),
\end{eqnarray}
where $\tau^{\ast}=\tau(b\tau)^{\mu-1}\left[1+(b\tau)^{-\rho}\right]^{\gamma}$, and it has a dimension of time $[\tau^{\ast}]=[\tau]=\mathrm{s}$. Graphical representation of the waiting time PDF is given in Fig.~\ref{fig_w2}. The influence of the tempering parameter $b$ on the behavior of the waiting time PDF is clearly demonstrated in Fig. \ref{fig_w2b}.

At the end, as a further generalization, we consider the distribution of the jump length of the form, in Fourier space,
\begin{equation}\label{jumplength}
\tilde{\lambda}(k)=1-(\sigma |k|)^{\alpha_1}[1+(\sigma |k|)^{-\rho_2}]^{\alpha_2},
\end{equation}
where $1<\alpha_1<2$ and $1<\alpha_1-\alpha_2\rho_2<2$. Combining (\ref{CTRW general}), (\ref{psi(s) prabhakar}), and (\ref{jumplength}), we arrive at
\begin{eqnarray}\label{general PDF}
\tilde{\hat{W}}(k,s)\nonumber=\frac{s^{-1}(s\tau)^\mu(1+(s\tau)^{-\rho})^\gamma\tilde{W}_{0}(k)}{(s\tau)^\mu(1+(s\tau)^{-\rho})^\gamma+(\sigma |k|)^{\alpha_1}(1+(\sigma |k|)^{-\rho_2})^{\alpha_2}}.
\end{eqnarray}
Note that for $\gamma=0$ and $\alpha_2=0$,  we recover the result for the L\'{e}vy flights with L\'{e}vy distribution of the jump length $\tilde{\lambda}(k)=e^{-\sigma^{\alpha_1}|k|^{\alpha_1}}\simeq1-\sigma^{\alpha_1}|k|^{\alpha_1}$ \cite{report}.

\section{Non-negativity of solution: Subordination approach, and Stochastic Representation}

Next we find the PDF which subordinates the diffusion processes, governed by the diffusion equation with Prabhakar time fractional derivative (\ref{diff eq}), from the time scale $t$ (physical time) to the Wiener processes on a time scale $u$ (operational time). In such a scheme the PDF $W(x,t)$ of a given random process $x(t)$ can be represented as \cite{barkai_pre2001,mmnp,mark}
\begin{equation}\label{wienersub}
W(x,t)=\int_0^{\infty}P(x,u)h(u,t)\,du,
\end{equation}
where
\begin{equation}\label{Wiener PDF}
P(x,u)=\frac{1}{\sqrt{4\pi\mathcal{K}_{1}u}}\exp\left(-\frac{x^2}{4\mathcal{K}_{1}u}\right),
\end{equation}
is the famed Gaussian PDF, i.e., the PDF of the Wiener process, and $h(u,t)$ is a PDF \emph{subordinating\/} the random process $x(t)$ to the Wiener process, and $\mathcal{K}_{1}$ is the diffusion coefficient with $\left[\mathcal{K}_{1}\right]=\mathrm{m}^{2}\mathrm{s}^{-1}$. Eq. (\ref{CTRW}) can be rewritten in the form
\begin{eqnarray}\label{PDF FL p1}
\tilde{\hat{W}}(k,s)=\int_{0}^{\infty}e^{-u\mathcal{K}_{1}k^{2}}\hat{h}(u,s)\,du,
\end{eqnarray}
where
\begin{eqnarray}\label{h1(u,s)}
\hat{h}(u,s)=s^{\mu-1}\left[1+(s\tau)^{-\rho}\right]^{\gamma}e^{-us^{\mu}\left[1+(s\tau)^{-\rho}\right]^{\gamma}}\nonumber =-\frac{\partial}{\partial u}\frac{1}{s}\hat{L}(s,u)
\end{eqnarray}
and
\begin{eqnarray}\label{L1(s,u)}
\hat{L}(s,u)=e^{-us^{\mu}\left[1+(s\tau)^{-\rho}\right]^{\gamma}}.
\end{eqnarray}
Taking inverse Fourier-Laplace transform of (\ref{PDF FL p1}) leads to
\begin{eqnarray}\label{PDF p1 general}
W(x,t)=\int_{0}^{\infty}\frac{1}{\sqrt{4\pi\mathcal{K}_{1}u}}e^{-\frac{x^{2}}{4\mathcal{K}_{1}u}}h(u,t)\,du,
\end{eqnarray}
which means that the PDF $h(u,t)$ subordinates the random processes governed by Eq.~(\ref{diff eq}) to the Wiener process by using the operational time $u$. From (\ref{PDF p1 general}), it can be seen that $W(x,t)$ is non-negative if $h(u,t)$ is non-negative, i.e., if $\hat{h}(u,s)$ is completely monotone function in respect to $s$. The function $\hat{h}(u,s)$ is completely monotone if both functions $s^{\mu-1}\left[1+(s\tau)^{-\rho}\right]^{\gamma}$ and $e^{-us^{\mu}\left[1+(s\tau)^{-\rho}\right]^{\gamma}}$ are completely monotone, and the function $e^{-us^{\mu}\left[1+(s\tau)^{-\rho}\right]^{\gamma}}$ is completely monotone if $s^{\mu}\left[1+(s\tau)^{-\rho}\right]^{\gamma}$ is a Bernstein function \cite{book_bernstein}\footnote{The function $e^{-uf(s)}$ is completely monotone if $f(s)$ is a Bernstein function.}. It can be easily verified that these conditions are satisfied if $0<\mu/\gamma$ and $0<\mu/\gamma-\rho<1$. Therefore, for these parameter constraints the PDF $W(x,t)$ is non-negative.

Following the procedure in \cite{marcin,marcin2,orzel}, we can construct a stochastic process $x(t)$, the PDF of which obeys the diffusion equation with Prabhakar derivative (\ref{diff eq}), and which can be represented as a rescaled Brownian motion $B(u)$ subordinated by an inverse L\'evy-stable subordinator $\mathcal{S}(t)$, independent of $B(u)$. Therefore, the stochastic process can be given by
\begin{eqnarray}\label{stochastic}
x(t)=\sqrt{2\mathcal{K}_{\mu}}B\left[\mathcal{S}(t)\right],
\end{eqnarray}
where the operational time is given by $\mathcal{S}(t)=\inf\left\{u>0: \mathcal{T}(u)>t\right\}$, and $\mathcal{T}(u)$ is an infinite divisible process, i.e., a strictly increasing L\'evy motion with $\left\langle e^{-s\mathcal{T}(u)}\right\rangle=e^{-u\hat{\Psi}(s)}$, and $\hat{\Psi}(s)=s^{\mu}\left[1+(s\tau)^{-\rho}\right]^{\gamma}$ is the L\'{e}vy exponent. To ensure  that the given process is well defined, the L\'evy exponent should belong to the class of Bernstein functions \cite{marcin,orzel}. As we show before, the function $\hat{\Psi}(s)=s^{\mu}\left[1+(s\tau)^{-\rho}\right]^{\gamma}$ is a Bernstein function for $0<\mu/\gamma<1$ and $0<\mu/\gamma-\rho<1$, so the stochastic process is well defined and its PDF satisfies the diffusion equation (\ref{diff eq}). Following the rigorous procedure given in \cite{marcin2} (see also \cite{polonica,physica_a}), from the infinitely divisible distribution $\mathcal{T}(u)$ representing the waiting time PDF in the CTRW model, one can find the corresponding diffusion equation of the form
\begin{eqnarray}\label{eta memory eq}
\frac{\partial}{\partial t}W(x,t)=\mathcal{K}_{\mu}\frac{d}{dt}\int_{0}^{t}\eta(t-t')\frac{\partial^2}{\partial x^{2}}W(x,t')\,dt',
\end{eqnarray}
where $\hat{\eta}(s)=\frac{1}{\hat{\Psi}(s)}=\frac{s^{-\mu+\rho\gamma}}{\left(s^{\rho}+\tau^{-\rho}\right)^{\gamma}}$. By inverse Laplace transform, for the memory kernel $\eta(t)$ we find
\begin{eqnarray}\label{eta exact}
\eta(t)=t^{\mu-1}E_{\rho,\mu}^{\gamma}\left(-\left[\frac{t}{\tau}\right]^{\rho}\right).
\end{eqnarray}
Therefore, the integro-differential equation (\ref{eta memory eq}) with memory kernel (\ref{eta exact}) has same solution with Eq.~(\ref{diff eq}) with Prabhakar derivative, which can be shown following the analysis in \cite{fcaa_2018}. Let us show this. In \cite{fcaa_2018} it is shown that Eq.~(\ref{eta memory eq}) has same solution with the one of the following equation
\begin{eqnarray}\label{gamma memory eq}
\int_{0}^{t}\gamma(t-t')\frac{\partial}{\partial t}W(x,t')\,dt'=\mathcal{K}_{\mu}\frac{\partial^2}{\partial x^{2}}W(x,t),
\end{eqnarray}
where the memory kernels $\gamma(t)$ and $\eta(t)$ in the Laplace space are connected as $\hat{\gamma}(s)\rightarrow1/[s\hat{\eta}(s)]$. Therefore, we have
\begin{eqnarray}\label{gamma exact}
\gamma(t)=\mathcal{L}^{-1}\left[\frac{1}{s\frac{s^{-\mu+\rho\gamma}}{\left(s^{\rho}+\tau^{-\rho}\right)^{\gamma}}}\right]=t^{-\mu}E_{\rho,1-\mu}^{-\gamma}\left(-\left[\frac{t}{\tau}\right]^{\rho}\right).
\end{eqnarray}
By substitution of the memory kernel (\ref{gamma exact}) in Eq.~(\ref{gamma memory eq}) we obtain Eq.~(\ref{diff eq}), i.e., the fractional diffusion equation with the Prabhakar derivative. We note that for $\gamma=0$ from Eq.~(\ref{eta memory eq}) one obtains the mono-fractional diffusion equation with R-L fractional derivative of order $1-\mu$ from the right hand side of the equation, since $\eta(t)=t^{\mu-1}/\Gamma(\mu)$. This equation is equivalent to Eq.~(\ref{gamma memory eq}) with $\gamma=0$ for which $\gamma(t)=t^{-\mu}/\Gamma(1-\mu)$, i.e., diffusion equation with Caputo fractional derivative of order $\mu$ from the left hand side of the equation.

\section{MSD, fractional moments, and multi-scaling}

Now, we calculate the MSD of the process described by the time fractional diffusion equation (\ref{diff eq}). We use the expression in Fourier-Laplace space, for $W_{0}(k)=1$,
\begin{equation}\label{msd general k=0}
\left\langle x^2(t)\right\rangle=\mathcal{L}^{-1}\left.\left[-\frac{\partial^2}{\partial k^2}\tilde{\hat{W}}(k,s)\right]\right|_{k=0}.
\end{equation}
From Eq.~(\ref{CTRW}), there exists
\begin{eqnarray}\label{MSD F=0}
\left\langle x^2(t)\right\rangle =2\mathcal{K}_{\mu}\mathcal{L}^{-1}\left[\frac{s^{-\mu-1}}{\left[1+(s\tau)^{-\rho}\right]^{\gamma}}\right]\nonumber =2\mathcal{K}_{\mu}t^{\mu}E_{\rho,\mu+1}^{\gamma}\left(-\left[\frac{t}{\tau}\right]^{\rho}\right),
\end{eqnarray}
from where for the short time limit we find
\begin{eqnarray}\label{MSD F=0 short}
\left\langle x^2(t)\right\rangle=2\mathcal{K}_{\mu}\frac{t^{\mu}}{\Gamma(\mu+1)},
\end{eqnarray}
and
\begin{eqnarray}\label{MSD F=0 long}
\left\langle x^2(t)\right\rangle=2\mathcal{K}_{\mu}\tau^{\mu}\frac{(t/\tau)^{\mu-\rho\gamma}}{\Gamma(\mu-\rho\gamma+1)}
\end{eqnarray}
in the long time limit, which means that decelerating subdiffusion exists in the system described by the diffusion equation with Prabhakar time fractional derivative. Here we note that the same result for the MSD as the one given by Eq.~(\ref{MSD F=0}) can also be obtained from Eq.~(\ref{msd general k=0}) if instead of long wave length approximation $\tilde{\lambda}(k)\simeq1-\sigma^{2}k^{2}$ one uses the exact jump length PDF $\tilde{\lambda}(k)=e^{-\sigma^{2}k^{2}}$ in the CTRW model (\ref{CTRW general}).

\begin{figure}\centering{\resizebox{0.75\textwidth}{!}{\includegraphics{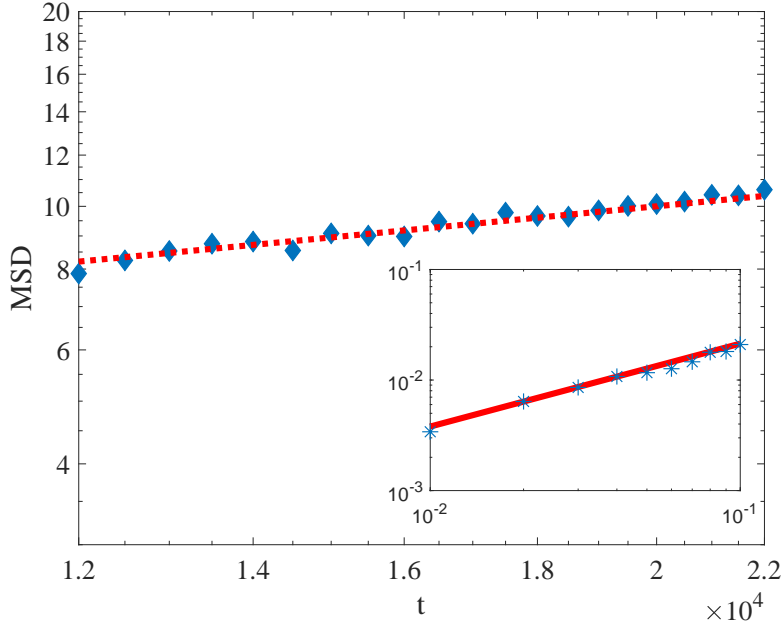}} \caption {Numerical simulations of the MSD by sampling $4\times10^4$ realizations in log-log scale. The dots are the simulation results by choosing the parameters of the waiting time distribution $\rho=7/16$, $\mu=3/4$, $\gamma=5/6$, $\tau=20$ and the jump length distribution as Gaussian, while red dotted (with the slope of $\mu-\rho\gamma=37/96\approx0.385$) and real lines (with the slope of $\mu=3/4=0.75$) are the theoretical results of MSD in long time and short time limit as shown in the inner figure respectively, being confirmed by the simulations.}\label{fig_MSD}}
\end{figure}

These results are verified by the numerical simulations shown in Fig. \ref{fig_MSD}. From the results shown in the above equations, we can also see the role of the parameter $\tau$, which controls the speed of transition, in fact, the same conclusion can also be drawn from Eq. (\ref{psi(s) prabhakar}). Specifically, the larger $\tau$ is, the slower transition will be, and vice versa. We note that the  decelerating subdiffusion has also been observed in the distributed order diffusion equations \cite{chechkin_1}.

%\begin{figure}\centering{\resizebox{0.75\textwidth}{!}{\includegraphics{MSD_transition.eps}} \caption {Numerical simulations of the MSD by sampling $4\times10^4$ realizations in log-log scale. The blue dots are the simulation results by choosing the parameters of the waiting time distribution $\rho=7/16$, $\mu=3/4$, $\gamma=5/6$, $\tau=10$ and the jump length distribution as Gaussian, while red lines are the theoretical results, being confirmed by the simulations.}\label{fig_MSD}}
%\end{figure}

As to the tempered time fractional diffusion equation (\ref{tempered diff eq}), for the MSD, there exists
\begin{eqnarray}\label{tempered MSD F=0}
\left\langle x^2(t)\right\rangle=2\mathcal{K}_{\mu}\,{^{\mathrm{RL}}}I_{0+}^{2}\left(e^{-bt}t^{\mu-2}E_{\rho,\mu-1}^{\gamma}\left(-\left[\frac{t}{\tau}\right]^{\rho}\right)\right),
\end{eqnarray}
where ${^{\mathrm{RL}}}I_{0+}^{\alpha}$ is the R-L fractional integral (\ref{RL integral}). Then, for the short time limit, it is obtained that $\left\langle x^{2}(t)\right\rangle\simeq t^{\mu}$, and for the long time limit a normal diffusion $\left\langle x^{2}(t)\right\rangle\simeq t$ appears. This means that the accelerating diffusion, from subdiffusion to normal diffusion, exists in the system. The normal diffusion in the long time limit appears due to the exponential tempering in the regularized Prabhakar derivative. These results are also in accordance with the results for the waiting time PDF, which in the short time limit has a behavior as the one without tempering, and in the long time limit it is an exponential waiting time PDF for Brownian motion. From (\ref{tempered MSD F=0}), in case of no tempering ($b=0$) we recover the result (\ref{MSD F=0}). Same situation of accelerated diffusion -- from subdiffusion to normal diffusion -- has been observed in the generalized Langevin equation with tempered regularized Prabhakar derivative \cite{sandev_mathematics2017}.

Next we calculate the fractional order moments $\left\langle|x(t)|^q\right\rangle$ of the diffusion equation (\ref{diff eq}), given by
\begin{eqnarray}\label{laplace Mq}
\mathcal{L}\left[\left\langle |x(t)|^q\right\rangle\right]=\int_{-\infty}^{\infty}|x|^{q}\hat{W}(x,s)\,dx.
\end{eqnarray}
Using $y=(s^{\mu/2}\left(1+(s\tau)^{-\rho}\right)^{\gamma/2}|x|$, from the PDF in the Laplace space, we obtain
\begin{eqnarray}\label{Mq exact}
\left\langle |x(t)|^q\right\rangle =\Gamma(q+1)\left(\mathcal{K}_{\mu}t^{\mu}\right)^{q/2}E_{\rho,\mu q/2+1}^{\gamma q/2}\left(-\left[\frac{t}{\tau}\right]^{\rho}\right).
\end{eqnarray}
Therefore, the short time limit $t/\tau\ll1$ yields
\begin{eqnarray}\label{Mq exact short}
\left\langle |x(t)|^q\right\rangle \simeq \Gamma(q+1)\frac{\left(\mathcal{K}_{\mu}t^{\mu}\right)^{q/2}}{\Gamma(1+\mu q/2)}\nonumber \left[1-\frac{\gamma q}{2}\frac{\Gamma(1+\mu q/2)}{\Gamma\left(1+\rho+\mu q/2\right)}(t/\tau)^{\rho}\right].
\end{eqnarray}
Here it can be noted that the leading term in the short time limit $t/\tau\ll1$ (i.e., $s\tau\gg1$) corresponds to the one obtained from the fractional Fokker-Planck equation where the fractional derivative is of order $\mu$. This can be seen from the general expression for the waiting time PDF (\ref{psi(s) prabhakar}) where for $s\tau\gg1$ it behaves as $\psi(s)\simeq\frac{1}{1+(s\tau)^{\mu}}$. This is exactly the same waiting time PDF for mono-fractional diffusion equation with anomalous diffusion exponent $\mu$. The second correction term in (\ref{Mq exact short}) appears as a result of Taylor expansion of the waiting time PDF (\ref{psi(s) prabhakar}) for $s\tau\gg1$, which has a form $\psi(s)\simeq\frac{1}{1+(s\tau)^{\mu}[1+\gamma(s\tau)^{-\rho}]}$. On the contrary, the long time limit behaves as
\begin{eqnarray}\label{Mq exact long}
\left\langle |x(t)|^q\right\rangle &\simeq \Gamma(q+1)\left(\mathcal{K}_{\mu}\tau^{\mu}\right)^{q/2}\frac{(t/\tau)^{(\mu-\rho\gamma)q/2}}{\Gamma(1+(\mu-\rho\gamma)q/2)}\nonumber \\&\times \left[1-\frac{\gamma q}{2}\frac{\Gamma(1+(\mu-\rho\gamma)q/2)}{\Gamma\left(1-\rho+(\mu-\rho\gamma)q/2\right)}(t/\tau)^{-\rho}\right].
\end{eqnarray}
The leading term in (\ref{Mq exact long}) corresponds to the one obtained from the fractional Fokker-Planck equation where the fractional derivative is of order $\mu-\rho\gamma$ since for the long time limit ($s\tau\ll1$) the waiting time PDF (\ref{psi(s) prabhakar}) behaves as $\psi(s)\simeq\frac{1}{1+(s\tau)^{\mu-\rho\gamma}}$. The second correction term in (\ref{Mq exact long}) is due to the complexity of the waiting time PDF. It is obtained as a result of Taylor expansion of the waiting time PDF (\ref{psi(s) prabhakar}) for $s\tau\ll1$, which has a form $\hat{\psi}(s)\simeq\frac{1}{1+(s\tau)^{\mu-\rho\gamma}[1+\gamma(s\tau)^{\rho}]}$. From here we conclude that the fractional order moments exhibit the scaling behavior
\begin{equation}\label{moments multifractal general}
\left\langle |x(t)|^q\right\rangle=C(q)t^{\nu(\mu,q,\gamma,t)},
\end{equation}
where $\nu(\mu,q,\gamma,t)$ is called the multi-scaling exponent. Such multi-scaling behavior exhibits, for example, the distributed order fractional diffusion equations \cite{sandev_pre2015,sandev_fcaa2015}. This is a more general result than a self-affine behavior $\left\langle |x(t)|^q\right\rangle=C(q)t^{qH}$ \cite{mandelbrot}, where $H>0$ is the Hurst exponent; for ordinary Brownian motion, $H=1/2$; for fractional Brownian motion, $0<H<1$; and $H=1/\alpha$ for L\'{e}vy flights as long as $q$ is smaller than the stable index $\alpha$, and for subdiffusive CTRW processes with scale-free, power-law waiting time PDF. Dynamics governed by the fractional diffusion equation with Caputo time fractional derivative belongs to the class of fractal or self-affine processes. The exponent $\nu\propto q$ has a linear dependence on the fractional order $q$. When $\left\langle |x(t)|^q\right\rangle=C(q)t^{\nu(q)}$, where $\nu(q)$ is a given nonlinear function, we call it a multi-fractal or multi-affine process \cite{mandelbrot}. In \cite{sandev_pre2015,sandev_fcaa2015} we show that the distributed order diffusion equations yield $q$-th moment of form (\ref{moments multifractal general}), as well.

Few words regarding the fractional moments are in order. The calculated fractional moments (\ref{Mq exact}) are exact and related to the fundamental solution of the fractional diffusion equation (\ref{diff eq}), which means that they correspond to the long wavelength approximation in the CTRW model. As we discussed before, for the long time approximation the MSD got from the fractional diffusion equation (\ref{diff eq}) is the same as the one obtained from the CTRW model with the exact Gaussian jump length PDF
\begin{eqnarray}\label{lambda exact}
\tilde{\lambda}(k)=e^{-\sigma^{2}k^{2}}&=1-\sigma^{2}k^{2}+\frac{\sigma^{4}k^{4}}{2}-\frac{\sigma^{6}k^{6}}{3!}+\dots \nonumber\\& =1-\frac{m_{2}k^{2}}{2}+\frac{m_{4}k^{4}}{4!}-\frac{m_6k^{6}}{6!}+\dots,
\end{eqnarray}
where $m_{2}=\Sigma^{2}=2\sigma^{2}$, $m_4=12\sigma^4$, $m_6=120\sigma^6$, i.e., $m_{2n}=2\sigma^{2n}$, $n=1,2,\dots$ are finite moments. If we calculate the fourth moment from the CTRW model (\ref{CTRW general}) for jump length PDF (\ref{lambda exact}) it can be found that the fourth moment depends not only on $m_2$ but also on $m_4$,
\begin{eqnarray}\label{M4 CTRW}
\left\langle x^{4}(t)\right\rangle=6\,m_2\left(\frac{t}{\tau}\right)^{2\mu}E_{\rho,2\mu+1}^{2\gamma}\left(-\left[\frac{t}{\tau}\right]^{\rho}\right)\nonumber+m_4\left(\frac{t}{\tau}\right)^{\mu}E_{\rho,\mu+1}^{\gamma}\left(-\left[\frac{t}{\tau}\right]^{\rho}\right).
\end{eqnarray}
From here in the long time limit, by using the asymptotic expansion formula for three parameter M-L function with large arguments, we find
\begin{eqnarray}\label{M4 exact long CTRW}
\left\langle x^4(t)\right\rangle \simeq &6m_{2}\frac{(t/\tau)^{2(\mu-\rho\gamma)}}{\Gamma(1+2(\mu-\rho\gamma))}\nonumber \left[1-2\gamma\frac{\Gamma(1+2(\mu-\rho\gamma))}{\Gamma\left(1-\rho+2(\mu-\rho\gamma)\right)}(t/\tau)^{-\rho}\right]\nonumber\\&+m_{4}\frac{(t/\tau)^{\mu-\rho\gamma}}{\Gamma(1+(\mu-\rho\gamma))}\nonumber \left[1-\gamma\frac{\Gamma(1+(\mu-\rho\gamma))}{\Gamma\left(1-\rho+(\mu-\rho\gamma)\right)}(t/\tau)^{-\rho}\right].
\end{eqnarray}
From here we find that a dominant term is the one with $m_2$, i.e., $\left\langle x^4(t)\right\rangle\simeq 6m_{2}\frac{(t/\tau)^{2(\mu-\rho\gamma)}}{\Gamma(1+2(\mu-\rho\gamma))}$, which is the same as the first term in (\ref{Mq exact long}) with $q=4$. Difference in the behavior of the fourth moment obtained from Eq.~(\ref{diff eq}), which is derived from the CTRW theory in the long wavelength approximation $\tilde{\lambda}(k)=1-\sigma^{2}k^{2}$, and from the CTRW model (Eq.~(\ref{M4 CTRW})) appears in the short time limit. In fact, from (\ref{M4 CTRW}) in the short time limit it can be found that the term with $m_4$ is dominant, i.e.,
\begin{eqnarray}
\left\langle x^{4}(t)\right\rangle\simeq&\frac{m_{4}}{\tau^{\mu}}\frac{(t/\tau)^{\mu}}{\Gamma(1+\mu)}\left[1-\gamma\frac{\Gamma(1+\mu)}{\Gamma(1+\mu+\rho)}(t/\tau)^{\rho}\right]\nonumber\\&+6m_{2}\frac{(t/\tau)^{2\mu}}{\Gamma(1+2\mu)}\left[1-2\gamma\frac{\Gamma(1+2\mu)}{\Gamma(1+2(\mu+\rho))}(t/\tau)^{\rho}\right].
\end{eqnarray}
This behavior is different from the result (\ref{Mq exact short}), obtained from Eq.~(\ref{diff eq}) for the long wavelength approximation, where all the higher order moments are set to be zero ($m_4=\dots=m_{2j}=0$). This point has been discussed by Barkai in \cite{barkai_cp2002} in detail for the case of mono-fractional diffusion equation (being also discussed in \cite{barkai2}), which is the case with $\gamma=0$ in our analysis.

\begin{figure}
\centering{\resizebox{0.75\textwidth}{!}{\includegraphics{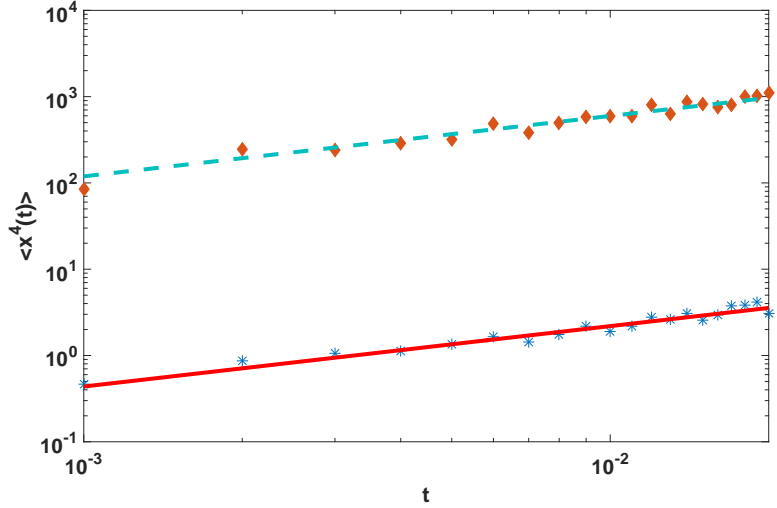}}}
\caption{Numerical simulations of $\big<x^4(t)\big>$ for short time in log-log scale with $2\times10^4$ particles. The parameters of waiting time distribution in Eq. (\ref{psi(t)_prabhakar short}) are taken as $\tau=1$, $\mu=0.7$, and $\gamma=0$. The distributions of jump length are taken as normal ones with mean zero and variances, respectively, as 4 and 64. In the case that the variance is 4, the simulation results are with dots and the real line is the theoretical result with the slope of $0.7$. For the case that the variance is 64, the simulation results are with diamonds and the dotted line is the theoretical results with the slope of 0.7 again.}\label{fig_m4}
\end{figure}

According to the analysis above, we make the numerical simulations of $\big<x^4(t)\big>$ for the short time. According to Eq. (\ref{psi(t)_prabhakar short}), we generate the waiting time random variables obeying $\psi(t)\simeq\frac{1}{\tau}\frac{(t/\tau)^{\mu-1}}{\Gamma(\mu)}$, and the simulation results are given in Fig. \ref{fig_m4}. From the simulations, one can conclude that for the short time $\big<x^4(t)\big>$ behaves as $t^{\mu}$ and the variances of the normal distribution that jump length random variables follow also affect $\big<x^4(t)\big>$ (but the slope does not change).

Then we consider the fractional moment of Eq. (\ref{general PDF}). When $t$ is short enough, that is, $s$ and $|k|$ are very large, we have
\begin{equation*}
\tilde{\hat{W}}(k,s)\sim \frac{1}{s}\frac{(s\tau)^\mu}{(s\tau)^\mu+(\sigma |k|)^{\alpha_1}}\tilde{W}_{0}(k).
\end{equation*}
If we rewrite this equation as
\begin{equation*}
s^{\mu}\tilde{\hat{W}}(k,s)-s^{\mu-1}\tilde{W}_{0}(k)=-\frac{{\sigma}^{\alpha_1}}{\tau^{\mu}}|k|^{\alpha_1}\tilde{\hat{W}}(k,s),
\end{equation*}
by inverse Fourier-Laplace transform, we arrive at the space-time fractional diffusion equation of the form
\begin{equation*}
{^{\mathrm{C}}}\mathcal{D}_{0+}^{\mu}W(x,t)=\frac{{\sigma}^{\alpha_1}}{\tau^{\mu}}\frac{\partial^{\alpha_1}}{\partial|x|^{\alpha_1}}W(x,t),
\end{equation*}
where $\frac{\partial^{\alpha}}{\partial|x|^{\alpha}}$ is the Riesz fractional derivative \cite{feller}\footnote{The Riesz fractional derivative of order $\alpha$ ($0<\alpha\leq2$) is given as a pseudo-differential operator with the Fourier symbol $-|k|^{\alpha}$, $k\in R$, i.e., $\frac{\partial^{\alpha}}{\partial|x|^{\alpha}}f(x)=\mathcal{F}^{-1}\left[-|k|^{\alpha}\tilde{F}(k)\right]$.}. Since the second moment does not exist, according to \cite{report}, we calculate the fractional moment
\begin{eqnarray}\label{general q moment short time}
\big<|x(t)|^q\big>=\frac{4\pi}{\alpha_1}\frac{\Gamma(1+q)\Gamma(1+q/\alpha_1)\Gamma(-q/\alpha_1)}{\Gamma(1+q/2)\Gamma(-q/2)}\frac{\left(\frac{\sigma^{\alpha_1}}{\tau^{\mu}}t^{\mu}\right)^{q/\alpha_1}}{\Gamma(1+\mu q/\alpha_1)},
\end{eqnarray}
where $0<q<\alpha_1<2$, and instead the MSD we calculate $\big<|x(t)|^q\big>^{2/q}\simeq t^{\frac{2\mu}{\alpha_1}}$. Therefore, there exists a competition between long rests and long jumps depending on the values of parameters. For $\frac{2\mu}{\alpha_1}<1$, one observes subdiffusion and for $\frac{2\mu}{\alpha_1}>1$ -- superdiffusion.

On the other hand, we consider the long time limit. Let $s$ and $|k|$ tend to 0, then
\begin{equation*}
\tilde{\hat{W}}(k,s)\sim \frac{1}{s}\frac{(s \tau)^{-\rho \gamma+\mu}}{(s \tau)^{-\rho \gamma+\mu}+(\sigma|k|)^{-\rho_2\alpha_2+\alpha_1}}\tilde{W}_{0}(k).
\end{equation*}
In the same way as previous we find the space-time fractional diffusion equation
\begin{equation*}
{^{\mathrm{C}}}\mathcal{D}_{0+}^{\mu-\rho\gamma}W(x,t)=\frac{{\sigma}^{\alpha_1-\rho_2\alpha_2}}{\tau^{\mu-\rho\gamma}}\frac{\partial^{\alpha_1-\rho_2\alpha_2}}{\partial|k|^{\alpha_1-\rho_2\alpha_2}}W(x,t).
\end{equation*}
Similarly, it can be got that
\begin{eqnarray}\label{general q moment long time}
\big<|x(t)|^q\big>=&\frac{4\pi}{\alpha_1-\rho_2\alpha_2}\frac{\Gamma(1+q)\Gamma(1+q/(\alpha_1-\rho_2\alpha_2))\Gamma(-q/(\alpha_1-\rho_2\alpha_2))}{\Gamma(1+q/2)\Gamma(-q/2)}\nonumber\\
&\times\frac{\left(\frac{\sigma^{\alpha_1-\rho_2\alpha_2}}{\tau^{\mu-\rho\gamma}}t^{\mu-\rho\gamma}\right)^{q/(\alpha_1-\rho_2\alpha_2)}}{\Gamma\left(1+\frac{(\mu-\rho\gamma)q}{\alpha_1-\rho_2\alpha_2}\right)},
\end{eqnarray}
where $0<q<\alpha_1-\rho_2\alpha_2<2$, and then $\big<|x(t)|^q\big>^{2/q}\simeq t^{\frac{2(\mu-\rho\gamma)}{\alpha_1-\rho_2\alpha_2}}$. Thus the system exhibits subdiffusion if $\frac{2(\mu-\rho\gamma)}{\alpha_1-\rho_2\alpha_2}<1$, and superdiffusion if $\frac{2(\mu-\rho\gamma)}{\alpha_1-\rho_2\alpha_2}>1$.

If we choose $\gamma=0$ (Caputo time fractional derivative), the $q$-th moment transits from $t^{(q \mu)/\alpha_1}$ to $t^{(q\mu)/(\alpha_1-\rho_2\alpha_2)}$ as time increases. Thus the process in this case accelerates. By controlling the parameters $\mu$, $\rho$, $\gamma$, $\alpha_1$, $\alpha_2$ and $\rho_2$ we can achieve accelerating or decelerating transition, besides we can also achieve different types of transitions among superdiffusion, normal diffusion and subdiffusion. From (\ref{general q moment short time}) and (\ref{general q moment long time}) we conclude that the scenario observed in the system depends on the values of the ratios $\rho\gamma/\mu$ and $\rho_2\alpha_2/\alpha_1$.

From all obtained results one can conclude that the proposed model is very general and versatile, and can describe different subdiffusive, normal diffusive and superdiffusive processes, as well as processes with crossover from one to another diffusive regime. Such diffusive processes include anomalous diffusion in biological cells \cite{franosh,metzlerPCCP}, transient diffusion of telomeres in the nucleus of mammalian cells \cite{Bronstein}, transient diffusion in plasma membrane \cite{Murase} and for other nuclear bodies \cite{Saxton}. Crossover from subdiffusion to normal diffusion has been observed in complex viscoelastic systems, for example in lipid bilayer systems \cite{jeon}.

\section{FFPE with Prabhakar derivative}

The fractional Fokker-Planck equation was introduced in \cite{mebakla} to describe an anomalous subdiffusive behavior of a particle in an external nonlinear field close to thermal equilibrium, being derived from the CTRW theory in \cite{barkai2000}. Here, we further analyze the case where the diffusing test particle is confined in an external potential $V(x)$. Following the derivations in \cite{carmi}, we can obtain the equation
\begin{equation}\label{FFPE memory}
{^{\mathrm{C}}}\mathcal{D}_{\rho,-\nu,0+}^{\gamma,\mu}W(x,t)=\left[\frac{\partial}{\partial x}\frac{V'(x)}{m\eta_{\mu}}+\mathcal{K}_{\mu}\frac{\partial^2}{\partial x^2}\right]W(x,t),
\end{equation}
where $\nu=\tau^{-\rho}$, $m$ is the mass of the particle, $\mathcal{K}_{\mu}$ is the generalized diffusion coefficient,  and $\eta_{\mu}$ is the friction coefficients with physical dimension $\left[\eta_{\mu}\right]=\mathrm{s}^{\mu-2}$. Here we note that from the definition of the Prabhakar derivative (\ref{prabhakar derivative}) and by setting $\frac{\partial}{\partial t}W(x,t)=0$, the generalized Einstein-Stokes relation
$$\mathcal{K}_{\mu}=\frac{k_BT}{m\eta_{\mu}}$$ is obtained.

In the case of constant external force $F(x)=-\frac{\mathrm{d}V(x)}{\mathrm{d}x}=F\Theta(t)$ ($V(x)=-Fx$), where $\Theta(t)$ is the Heaviside step function, from the Fourier-Laplace transform of Eq.~(\ref{FFPE memory}), one gets
\begin{equation}\label{W_FLspace}
\tilde{\hat{W}}(k,s)=\frac{s^{\mu-1}\left[1+(s\tau)^{-\rho}\right]^{\gamma}}{s^{\mu}\left[1+(s\tau)^{-\rho}\right]^{\gamma}+\imath\frac{F}{m\eta_{\mu}}k+\mathcal{K}_{\mu}k^2}.
\end{equation}
Taking the inverse Fourier transform leads to the solution in the Laplace space
\begin{equation}\label{W_Lspace}
\hat{W}(x,s)=\hat{W}_{0}(x,s)\exp\left[-\frac{F}{2m\eta_{\mu}\mathcal{K}_{\mu}}x\right],
\end{equation}
where
\begin{eqnarray}\label{W_LspaceP}
\hat{W}_{0}(x,s)=\frac{\exp{\left[-\sqrt{\frac{s^{\mu}\left[1+(s\tau)^{-\rho}\right]^{\gamma}}{\mathcal{K}_{\mu}}+\left(\frac{F}{2m\eta_{\mu}\mathcal{K}_{\mu}}\right)^2}\,|x|\right]}}{\sqrt{\frac{s^{\mu}\left[1+(s\tau)^{-\rho}\right]^{\gamma}}{\mathcal{K}_{\mu}}+\left(\frac{F}{2m\eta_{\mu}\mathcal{K}_{\mu}}\right)^2}}\nonumber \frac{s^{\mu-1}\left[1+(s\tau)^{-\rho}\right]^{\gamma}}{2\mathcal{K}_{\mu}}.
\end{eqnarray}
In absence of external force field $F=0$, the PDF becomes
\begin{eqnarray}\label{PDF L diffusion like}
\hat{W}(x,s)=\frac{1}{2s}\sqrt{\frac{s^{\mu}\left[1+(s\tau)^{-\rho}\right]^{\gamma}}{\mathcal{K}_{\mu}}}{e^{-\sqrt{\frac{s^{\mu}\left[1+(s\tau)^{-\rho}\right]^{\gamma}}{\mathcal{K}_{\mu}}}|x|}}.
\end{eqnarray}

From the PDF (\ref{W_FLspace}) we can calculate the moments
$\left\langle x^n(t)\right\rangle=\mathcal{L}^{-1}\left.\left[i^n\frac{\partial^n}{\partial k^{n}}\tilde{\hat{W}}(k,s)\right]\right|_{k=0}$. Then, for the first moment, there exists
\begin{eqnarray}\label{meanF}
\left\langle x(t)\right\rangle_{F}=\frac{F}{m\eta_{\mu}}\mathcal{L}^{-1}\left[\frac{s^{-\mu-1}}{\left[1+(s\tau)^{-\rho}\right]^{\gamma}}\right]\nonumber= \frac{F}{m\eta_{\mu}}t^{\mu}E_{\rho,\mu+1}^{\gamma}\left(-\left[\frac{t}{\tau}\right]^{\rho}\right);
\end{eqnarray}
and for the second moment, we have
\begin{eqnarray}\label{MSD F}
\left\langle x^2(t)\right\rangle_{F}& =2\mathcal{K}_{\mu}\mathcal{L}^{-1}\left[\frac{s^{-\mu-1}}{\left[1+(s\tau)^{-\rho}\right]^{\gamma}}\right]+2\left(\frac{F}{m\eta_{\mu}}\right)^2\mathcal{L}^{-1}\left[\frac{s^{-2\mu-1}}{\left[1+(s\tau)^{-\rho}\right]^{2\gamma}}\right]\nonumber \\&=2\mathcal{K}_{\mu}t^{\mu}E_{\rho,\mu+1}^{\gamma}\left(-\left[\frac{t}{\tau}\right]^{\rho}\right)+2\left(\frac{F}{m\eta_{\mu}}\right)^2t^{2\mu}E_{\rho,2\mu+1}^{2\gamma}\left(-\left[\frac{t}{\tau}\right]^{\rho}\right).\nonumber\\
\end{eqnarray}
From (\ref{meanF}) and (\ref{MSD F}), we conclude that the second Einstein relation is satisfied, i.e., $$\left\langle x(t)\right\rangle_{F}=\frac{F}{2k_BT}\left\langle x^2(t)\right\rangle_0.$$

\subsection{Relaxation of modes}

By using the variable separation ansatz and taking $W(x,t)=X(x)T(t)$, Eq.~(\ref{FFPE memory}) yields
\begin{eqnarray}\label{memory FP eq T}
&{^{\mathrm{C}}}\mathcal{D}_{\rho,-\nu,0+}^{\gamma,\mu}T(t)=-\lambda T(t),\\
\label{memory FP eq X}
&\left[\frac{\partial}{\partial x}\frac{V'(x)}{m\eta_{\mu}}+\mathcal{K}_{\mu}\frac{\partial^2}{\partial x^2}\right]X(x)=-\lambda X(x),
\end{eqnarray}
where $\lambda$ is a separation constant. The solution of Eq.~(\ref{FFPE memory}) is given as $W(x,t)=\sum_nX_n(x)T_n(t)$, where $X_n(x)T_n(n)$ is the eigenfunction corresponding to the eigenvalue $\lambda_n$.

By employing Laplace transform of Eq.~(\ref{memory FP eq T}), we obtain the relaxation law
\begin{eqnarray}\label{memory schroedinger eq T solution}
T_n(t)=T_{n}(0)\sum_{j=0}^{\infty}(-\lambda_{n})^{j}t^{\mu j}E_{\rho,\mu j+1}^{\gamma j}\left(-\left[\frac{t}{\tau}\right]^{\rho}\right),
\end{eqnarray}
where $T_n(0)=\big<W_0(x),X_n(x)\big>$, the inner product of $W_0(x)$ and $X_n(x)$. From here, for the long time limit $t/\tau\gg1$, we find the power-law decay
\begin{eqnarray}\label{memory schroedinger eq T solution asymptotic}
T_n(t)\simeq\frac{T_n(0)}{\lambda_n\tau^{\mu}}\frac{(t/\tau)^{-\left(\mu-\rho\gamma\right)}}{\Gamma\left(1-\left(\mu-\rho\gamma\right)\right)}.
\end{eqnarray}
Note that for $\gamma=0$ we recover the result of mono-fractional diffusion equation
\begin{eqnarray*}
T_n(t)=T_{n}(0)\sum_{j=0}^{\infty}\frac{(-\lambda_{n})^{j}t^{\mu j}}{\Gamma(\mu j +1)}=T_{n}(0)E_{\mu}\left(-\lambda_{n}t^{\mu}\right).
\end{eqnarray*}

\subsection{Harmonic external potential}

\begin{figure}
  \centering
  \includegraphics[width=0.75\textwidth]{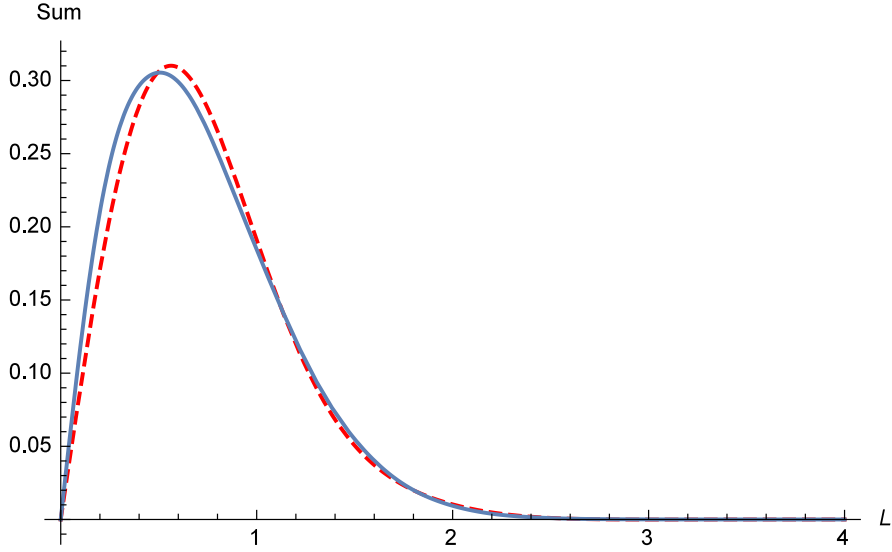}
  \caption{Sum over the first 10 and 100 terms of the second term of the infinite series in (\ref{surival}). By comparing the first 10 terms (red dash line) and the first 100 terms (blue solid line), it can be noted that the differences between these two sums can be neglected.}\label{fig2}
\end{figure}

The solution of the spatial eigenequation (\ref{memory FP eq X}) for the physically important case of an external harmonic potential $V(x)=\frac{1}{2}m\omega^2x^2$, where $\omega$ is a frequency, is given in terms of Hermite polynomials $H_n(z)$ \cite{erdelyi}
\begin{equation}
X_n(x)=\mathcal{C}_{n}H_n\left(\sqrt{\frac{m\omega^2}{2k_BT}}x\right)\exp\left(-\frac{m\omega^2}{2k_BT}x^2\right),
\end{equation}
where the eigenvalue spectrum (of the corresponding Sturm-Liouville problem) is given by $\lambda_n=n\frac{\omega^2}{\eta_{\mu}}$ for $n=0,1,2,...$, and $\mathcal{C}_{n}$ is the normalisation constant. From the normalisation condition $\big<X_n(x),X_n(x)\big>=1$, we obtain the solution of the form (compare Refs.~\cite{mebakla,report} for mono-fractional diffusion equation)
\begin{eqnarray}\label{relaxation}
W(x,t)=&\left(\frac{m\omega^2}{2\pi k_BT}\right)^{\frac{1}{2}}\sum_n\frac{1}{2^nn!}T_n(0)H_n\left(\sqrt{\frac{m\omega^2}{2k_BT}}x\right)\nonumber\\&\times\exp\left(-\frac{m\omega^2}{2k_BT}x^2\right)\sum_{j=0}^{\infty}\left(-\frac{n\omega^{2}}{\eta_{\mu}}\right)^{j}t^{\mu j}E_{\rho,\mu j+1}^{\gamma j}\left(-\left[\frac{t}{\tau}\right]^{\rho}\right).\nonumber\\
\end{eqnarray}
For $n=0$ it follows the Gaussian stationary solution $W(x,t)=\sqrt{\frac{m\omega^2}{2\pi k_BT}}\exp\left(-\frac{m\omega^2}{2k_BT}x^2\right)$. From \cite{dybiec}, we have the relation between the first passage time density $f(t)$ and the survival probability $S(t)$:
\begin{equation*}
F(t):=\int_{0}^{t}f(u)\,du=1-S(t),
\end{equation*}
where $S(t)=\int_{\Omega}W(x,t|x_0,0)\,dx$. That is $f(t)=-\frac{d}{dt}S(t)$. For the long time behavior, the sum over $j$ in (\ref{relaxation}) yields
\begin{equation*}
T_n(t)\sim T_n(0)\frac{\eta_\mu}{n \omega^2}\frac{\tau^{-\rho \gamma}t^{-(\mu-\rho \gamma)}}{\Gamma(1-\mu+\rho \gamma)}.
\end{equation*}
Thus
\begin{eqnarray*}
W(x,t|x_0,0)\sim&\sqrt{\frac{m\omega^2}{2\pi k_BT}}\exp\left(-\frac{m\omega^2}{2k_BT}x^2\right)+\left(\frac{m\omega^2}{2\pi k_BT}\right)^{\frac{1}{2}}\sum_{n=1}^{\infty}\frac{T_n(0)}{2^nn!}\nonumber\\
&\times H_n\left(\sqrt{\frac{m\omega^2}{2k_BT}}x\right)\exp\left(-\frac{m\omega^2}{2k_BT}x^2\right)\frac{\eta_\mu}{n \omega^2}\frac{\tau^{-\rho \gamma}t^{-(\mu-\rho \gamma)}}{ \Gamma(1-\mu+\rho \gamma)}.
\end{eqnarray*}
Here we only consider the case that the domain $\Omega$ is an interval $[-L,L]$. Then
\begin{eqnarray}\label{surival}
S(t)&\sim  \left(\frac{m\omega^2}{2\pi k_BT}\right)^{\frac{1}{2}}\int_{-L}^{L}\exp\left(-\frac{m\omega^2}{2k_BT}x^2\right)\,dx
+\left(\frac{m\omega^2}{2\pi k_BT}\right)^{\frac{1}{2}}\sum_{n=1}^{\infty}\frac{T_n(0)}{2^nn!}\nonumber\\
&\times\left[\int_{-L}^{L}H_n\left(\sqrt{\frac{m\omega^2}{2k_BT}}x\right)\exp\left(-\frac{m\omega^2}{2k_BT}x^2\right)\,dx\right]\frac{\eta_\mu\tau^{-\mu}}{n\omega^2}\frac{(t/\tau)^{-(\mu-\rho \gamma)}}{ \Gamma(1-\mu+\rho \gamma)}\nonumber\\&
={\rm erf}\left(\sqrt{\frac{m\omega^2}{2k_BT}}L\right)+\left(\frac{m\omega^2}{2\pi k_BT}\right)^{\frac{1}{2}}\sum_{n=1}^{\infty}\frac{T_n(0)}{2^nn!}\Bigg[\int_{-L}^{L}H_n\left(\sqrt{\frac{m\omega^2}{2k_BT}}x\right)\nonumber\\
&\times\exp\left(-\frac{m\omega^2}{2k_BT}x^2\right)\,dx\Bigg]\frac{\eta_\mu\tau^{-\mu}}{n\omega^2}\frac{(t/\tau)^{-(\mu-\rho \gamma)}}{ \Gamma(1-\mu+\rho \gamma)},
\end{eqnarray}
where ${\rm erf}(x)=\frac{1}{\sqrt{\pi}}\int_{-x}^{x}e^{-t^2}\,dt=\frac{2}{\sqrt{\pi}}\int_{0}^{x}e^{-t^2}\,dt$ is the error function \cite{erdelyi}. If we take the initial condition $W_0(x)=\delta(x)$, then $T_n(0)=H_n(0)$. And we assume $\frac{m\omega^2}{2k_BT}=1$. By the numerical calculations, we obtain that if $L\geqslant2$, the second term on the right hand side of (\ref{surival}) is $0$. Thus the survival probability asymptotically behaves as $S(t)\sim {\rm erf}(L)$. Then the density of the first passage time $f(t)=\delta(t)$. From this result, we can see the influence of external harmonic potential. Specifically, the particles are constrained in a small domain by the harmonic potential. On the other hand, we consider $0<L<2$. For this case we add the first 10 terms and the first 100 terms. We find the differences are very small. The numerical result of the sum over the first 100 terms is shown in Fig. \ref{fig2}.

Therefore, we only use the first 10 terms to approximate (\ref{surival})
\begin{eqnarray*}
S(t)\thickapprox & {\rm erf}(L)+\frac{\exp(-L^2)L}{115200\sqrt{\pi}}(104745 - 61270 L^2  + 18572 L^4 - 2328 L^6 \\
&+ 96 L^8)\frac{\eta_\mu}{ \omega^2}\frac{\tau^{-\rho \gamma}t^{-(\mu-\rho \gamma)}}{ \Gamma(1-\mu+\rho \gamma)}.
\end{eqnarray*}
Then the distribution of the first passage time is
\begin{eqnarray*}
f(t)\thickapprox & \frac{\exp(-L^2)L}{115200\sqrt{\pi}}(104745 - 61270 L^2 + 18572 L^4  - 2328 L^6 \\&
 + 96 L^8)\frac{\eta_\mu}{\tau^{\rho \gamma} \omega^2}\frac{\mu-\rho\gamma}{\Gamma(1-\mu+\gamma\rho)}t^{-1+\rho \gamma-\mu}.
\end{eqnarray*}

In case of the harmonic potential, we derive the corresponding differential equations for the first and second moments. Then, for the first moment, we get the following fractional equation with Prabhakar derivative
\begin{equation}\label{diff eq first moment harmonic}
{^{\mathrm{C}}}\mathcal{D}_{\rho,-\nu,0+}^{\gamma,\mu}\left\langle x(t)\right\rangle +\frac{\omega^2}{\eta_{\mu}}\left\langle x(t)\right\rangle=0.
\end{equation}
From here, by Laplace transform method, we find the relaxation law for the initial condition $x_0=\int_{-\infty}^{\infty}xW_0(x)\,dx$,
\begin{eqnarray}\label{mean relaxation law}
\left\langle x(t)\right\rangle=\sum_{j=0}^{\infty}\left(-\frac{\omega^{2}}{\eta_{\mu}}\right)^{j}t^{\mu j}E_{\rho,\mu j+1}^{\gamma j}\left(-\left[\frac{t}{\tau}\right]^{\rho}\right).
\end{eqnarray}
Taking $\gamma=0$ recovers the result for mono-fractional diffusion equation $\left\langle x(t)\right\rangle=x_0E_{\mu}\left(-\frac{\omega^2}{\eta_{\mu}}t^{\mu}\right)$ \cite{mebakla,report}, which in the long time limit shows the power-law scaling $\left\langle x(t)\right\rangle\simeq\frac{x_0\eta_{\mu}}{\omega^2}\frac{t^{-\mu}}{\Gamma(1-\mu)}$.

Respectively, for the second moment we obtain the following fractional equation with Prabhakar derivative
\begin{equation}\label{diff eq second moment harmonic}
{^{\mathrm{C}}}\mathcal{D}_{\rho,-\nu,0+}^{\gamma,\mu}\left\langle x^2(t)\right\rangle+2\frac{\omega^2}{\eta_{\mu}}\left\langle x^2(t)\right\rangle=2\mathcal{K}_{\mu},
\end{equation}
from which it follows that
\begin{eqnarray}
\left\langle x^2(t)\right\rangle =x_0^2\mathcal{L}^{-1}\left[\frac{s^{\mu-1}\left[1+(s\tau)^{-\rho}\right]^{\gamma}}{s^{\mu}\left[1+(s\tau)^{-\rho}\right]^{\gamma}+2\frac{\omega^2}{\eta_{\mu}}}\right]\mathcal{L}^{-1}\left[\frac{2\mathcal{K}_{\mu}}{s^{\mu}\left[1+(s\tau)^{-\rho}\right]^{\gamma}+2\frac{\omega^2}{\eta_{\mu}}}\right],\nonumber\\
\end{eqnarray}
or, equivalently
\begin{eqnarray}
\left\langle x^2(t)\right\rangle=x_{\mathrm{th}}^2+\left(x_0^2-x_{\mathrm{th}}^2
\right)\sum_{j=0}^{\infty}\left(-2\frac{\omega^{2}}{\eta_{\mu}}\right)^{j}t^{\mu j}E_{\rho,\mu j+1}^{\gamma j}\left(-\left[\frac{t}{\tau}\right]^{\rho}\right),
\end{eqnarray}
where $x_0=x(0)$ is the initial value of the position, and $x_{\mathrm{th}}^2=\frac{k_{B}T}{m\omega^2}$ is the stationary (thermal) value, being reached in the long time limit. Contrary to the case of Brownian motion where the second moment approaches the stationary value exponentially, in this case the second moment approaches the stationary value by Mittag-Leffler relaxation $$\left\langle x^2(t)\right\rangle\simeq x_{\mathrm{th}}^2+\left(x_0^2-x_{\mathrm{th}}^2\right)E_{\mu-\rho\gamma}\left(-\frac{2\omega^2}{\eta_\mu \tau^{-\mu}}\left[\frac{t}{\tau}\right]^{\mu-\rho\gamma}\right),$$ which turns to power-law in the long time limit, i.e.,
\begin{eqnarray*}
\left\langle x^2(t)\right\rangle\simeq x_{\mathrm{th}}^2+\left(x_0^2-x_{\mathrm{th}}^2\right)\frac{\eta_\mu \tau^{-\mu}}{2\omega^2}\frac{(t/\tau)^{-\mu+\rho\gamma}}{\Gamma(1-\mu+\rho\gamma)}.
\end{eqnarray*}

\section{Conclusion}

This paper focuses on building and analyzing the models, characterizing the transition of anomalous diffusion with different diffusion exponents. The analytical results for the waiting time PDF, MSD and fractional moments are obtained, showing the multi-scaling properties of the underlying stochastic processes. The non-negativity of the waiting time PDF and the solution of the equations, and the restrictions on parameters are carefully discussed. Both the equations with regularized Prabhakar derivative and tempered regularized Prabhakar derivative are analyzed, and the stochastic representation of the equation with regularized Prabhakar derivative is presented. We give the exact results for the relaxation of modes, mean displacement and MSD of the models. It is shown that the second moment has Mittag-Leffler approach to the thermal value.

\section*{Acknowledgements}
This work was supported by the National Natural Science Foundation of China under Grant No. 11671182, and the Fundamental Research Funds for the Central Universities under Grants No. lzujbky-2018-it60, No. lzujbky-2018-ot03, and No. lzujbky-2017-ot10. TS acknowledges funding from the Deutsche Forschungsgemeinschaft (DFG), project ME 1535/6-1 ``Random search processes, L\'evy flights, and random walks on complex networks". The authors are also thankful to Prof. Dr. Eli Barkai for the helpful discussion.

\end{document}